\documentclass{PoS}

\title{The spectrum of the light component of TeV cosmic rays measured with HAWC}

\ShortTitle{HAWC spectrum of TeV protons plus helium}

\author{
\speaker{J.C. Arteaga-Vel\'azquez} and J.D. \'Alvarez  
for the HAWC Collaboration\footnote{A complete list of authors is 
available at  https://www.hawc-observatory.org/collaboration/icrc2019.php
} \\
Instituto de F\'\i sica y Matem\'aticas, Universidad Michoacana, Morelia, Mexico\\
E-mail: \email{arteaga@ifm.umich.mx}       
}

\abstract{
We present a measurement of the energy spectrum of the light mass group of 
cosmic rays (protons and helium) with the High Altitude Water Cherenkov (HAWC) 
Observatory in the energy interval from  $10$ TeV to $200$ TeV. The spectrum 
covers the energy range between direct and indirect measurements, where 
precision data on the composition of cosmic rays are needed. The spectrum was 
constructed by applying an unfolding technique on a proton plus helium 
enriched sub-sample ($>90\%$ abundance) of cosmic ray air showers selected 
from the HAWC cosmic-ray data. The subset contains $3.8 \times 10^{9}$  air 
shower events with primary energies in the interval $1$ TeV to $300$ TeV 
and zenith angles less than $16.7$ degrees. Mass selection was performed 
using an analysis of the lateral age parameter of air shower events, based 
on both CORSIKA/QGSJET-II-03 simulations and event-by-event measurements 
on the lateral structure of the shower front and the primary energy of 
the showers.
}

\FullConference{36th International Cosmic Ray Conference -ICRC2019-\\
		July 24th - August 1st, 2019\\
		Madison, WI, U.S.A.}

\begin{document}

 \section{Introduction}
       \vspace{-0.5pc}
   The energy region from $10^{13} \, \mbox{eV}$ to $10^{15}  \, \mbox{eV}$ of the cosmic ray 
  energy spectrum has been barely explored as it is located at the frontier between the 
  direct and indirect detection techniques. However, it is of great interest due to 
  the possible presence of new structures in the spectra of all-particles and
  elemental mass groups of cosmic rays in this energy regime (see, for example,
  \cite{Stanev, nucleon, cream17, Argo15b}). To investigate the possible existence of such 
  structures, TeV cosmic-ray data with good precision and high statistics are required
  and HAWC can contribute to this task due to its capabilities.  HAWC is an 
  extensive air shower (EAS) observatory located at $4,100 \, \mbox{m}$ \textit{a.s.l.} 
  ($\sim 640 \, \mbox{g}/\mbox{cm}^2$ atmospheric depth) on the northern slope of the 
  volcano Sierra Negra in Mexico. The instrument is designed to detect $\gamma$ rays 
  from $100 \, \mbox{GeV}$ to $100 \, \mbox{TeV}$, but its altitude and physical 
  dimensions permit measurements of primary hadronic cosmic rays up to PeV energies.
  In this contribution, we will present a description of an unfolding analysis to 
  estimate the proton plus helium spectrum of cosmic rays in the energy interval 
  $10$ TeV to $200$ TeV with HAWC. We will also show the result and we will compare it
  with the measurements of other direct and indirect experiments.

  \section{Event reconstruction and simulation}
     \vspace{-0.5pc}
   HAWC is composed of $300$ water Cherenkov detectors (WCD), which are organized in 
  a compact and tight configuration on a flat surface of $22,000 \, \mbox{m}^2$. 
  Each WCD is made of a steel tank, $4.5$ m deep and $7.3$ m in diameter, filled with 
  $200, 000$ lt of water and $4$ upward-facing photomultipliers (PMTs) anchored 
  at the bottom of the tank. During the passage of an EAS, the shower particles
  produce Cherenkov light in the tanks, which produces voltage pulses at the PMTs. 
  The corresponding signals are then converted by a dedicated electronics 
  into an effective charged, $Q_{eff}$. The timing information and the signals of the 
  PMTs are then used as an input in a reconstruction software to estimate relevant 
  EAS observables of the event, such as its arrival direction, core position at 
  ground-level, lateral distribution of the deposited charge, lateral shower age 
  (hereafter  referred to as age) and primary energy \cite{HAWCcrab17, HAWCcrspec}.
  
 The age, $s$, is obtained event-by-event from a fit to the lateral charge 
 distribution measured by the PMTs at the shower plane with an NKG-like
 function \cite{NKG}:
 \begin{equation}
 f_{ch}(r) = A \cdot (r/r_0)^{s-3} \cdot (1 + r/r_0)^{s-4.5},
 \end{equation} 
 where $r$  is the radial distance to the shower axis, 
 $r_0 = 124.21 \, \mbox{m}$ is the Moliere radius and $A$ is a normalization 
 parameter.  This lateral distribution function gives a good description of 
 gamma-induced EAS \cite{Kelly}, and a reasonable description of 
 hadron-induced air showers detected with HAWC \cite{Morales}. 
 
  On the other hand, the primary energy of the event is estimated from a 
 maximum log-likelihood procedure, which computes and compares the  
 probabilities that the measured lateral distribution $Q_{eff} (r)$ (including PMTs with no signals)is
 produced by air showers of different energies. The algorithm makes use of 
 four-dimensional probability tables that are generated from proton-induced 
 EAS simulations and which are binned in primary energy, zenith angle, 
 $Q_{eff}$, and radial distance of the PMT to the shower core (for further 
 explanations see \cite{HAWCcrspec}).
 
  Air showers were simulated using CORSIKA v740\cite{Heck:1998vt} with 
 FLUKA \cite{Fluka} and QGSJet-II-03 \cite{qgsjet} as low-energy and 
 high-energy hadronic interaction models, respectively. As primary 
 nuclei, eight species were considered H, He, C, O, Ne, Mg, Si and Fe.
 They were generated with an $E^{-2}$ differential energy spectrum 
 for energies between $5 \, \mbox{GeV}$ and $3 \, \mbox{PeV}$, 
 zenith angles $\theta < 70^{\circ}$ and shower cores within an area 
 of $1 \, \mbox{km}$ of radius from the center of the array. On the 
 other hand, interactions 
 of the shower particles with the HAWC detector were simulated using a 
 code based on GEANT4 \cite{Geant4}. The output of the software has the 
 same format as the experimental data, which allows to reconstruct the
 MC events with the same code used for the real data. Finally,  an energy 
 weighting was introduced in the MC data sets to reproduce the best fit 
 spectra (with broken power-laws) to the direct measurements  from AMS 
 \cite{ams}, CREAM \cite{cream}, and PAMELA \cite{pamela}. Details
 of our nominal composition model are found in \cite{HAWCcrspec}.
 
  \section{Data selection}
     \vspace{-0.5pc}
    In order to reduce the influence of systematic uncertainties  in our analysis
  we have applied different selection cuts to our experimental and MC
  data samples. They were optimized 
  from studies with simulations. In particular, we only consider vertical 
  events with $\theta < 16.7^\circ$ that have successfully passed the reconstruction 
  algorithm, and have a minimum number of PMTs above threshold $N_{hit} = 75$.
  In addition, to reduce the uncertainty from the core position, we selected on-array 
  events with at least 60 PMTs activated within a radius of $40 \, \mbox{m}$
  from the EAS core. Besides, to decrease the bias in the reconstructed energy, 
  we also removed low energy events  from  the low efficiency region 
  ($\log_{10}(E/\mbox{GeV}) < 3.5$) using $f_{hit} \geq 0.3$, where $f_{hit}$ is the fraction 
  of hit PMTs at HAWC during the event  \cite{HAWCcrab17}. Finally, we constrained our 
  analysis only to data with energies $\log_{10}(E/\mbox{GeV}) < 5.5$. 
 
   For the present analysis we have used data from the HAWC DAQ period from June 11th, 2015
  up to November 28th, 2018, which contains almost $3 \times 10^{12}$ events. After selection
  cuts, we kept around $5.8 \times 10^{9}$ events, which corresponds to a total effective time
  of  $\Delta t_{eff} = 3.24 \, \mbox{yr}$ ($94 \%$ of livetime). According to MC simulations, at 
  energies $E \geq 10^{4} \, \mbox{GeV}$  the mean systematic uncertainties of the selected data 
  sample are $\delta R \leq 9 \, \mbox{m}$ for the shower core position, 
  $\delta \log_{10}(E/\mbox{GeV}) \leq 0.12$ for the 
  reconstructed energy, and $\delta \alpha \leq 0.3^{\circ}$ for the arrival direction 
  of the EAS. 
  
     \begin{figure}[!tp]
     \centering
     \footnotesize
     \includegraphics[width=75mm]{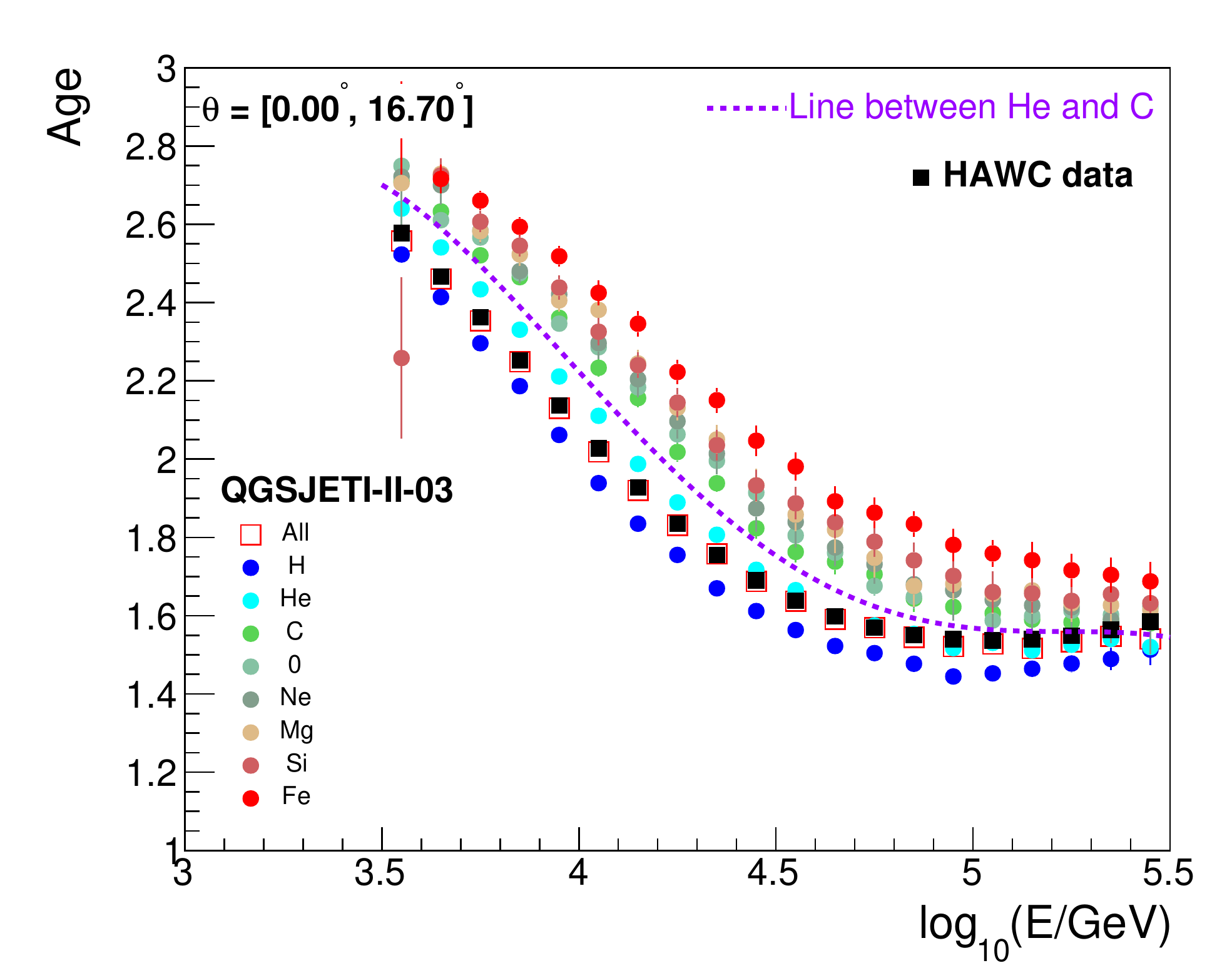} 
     \includegraphics[width=75mm]{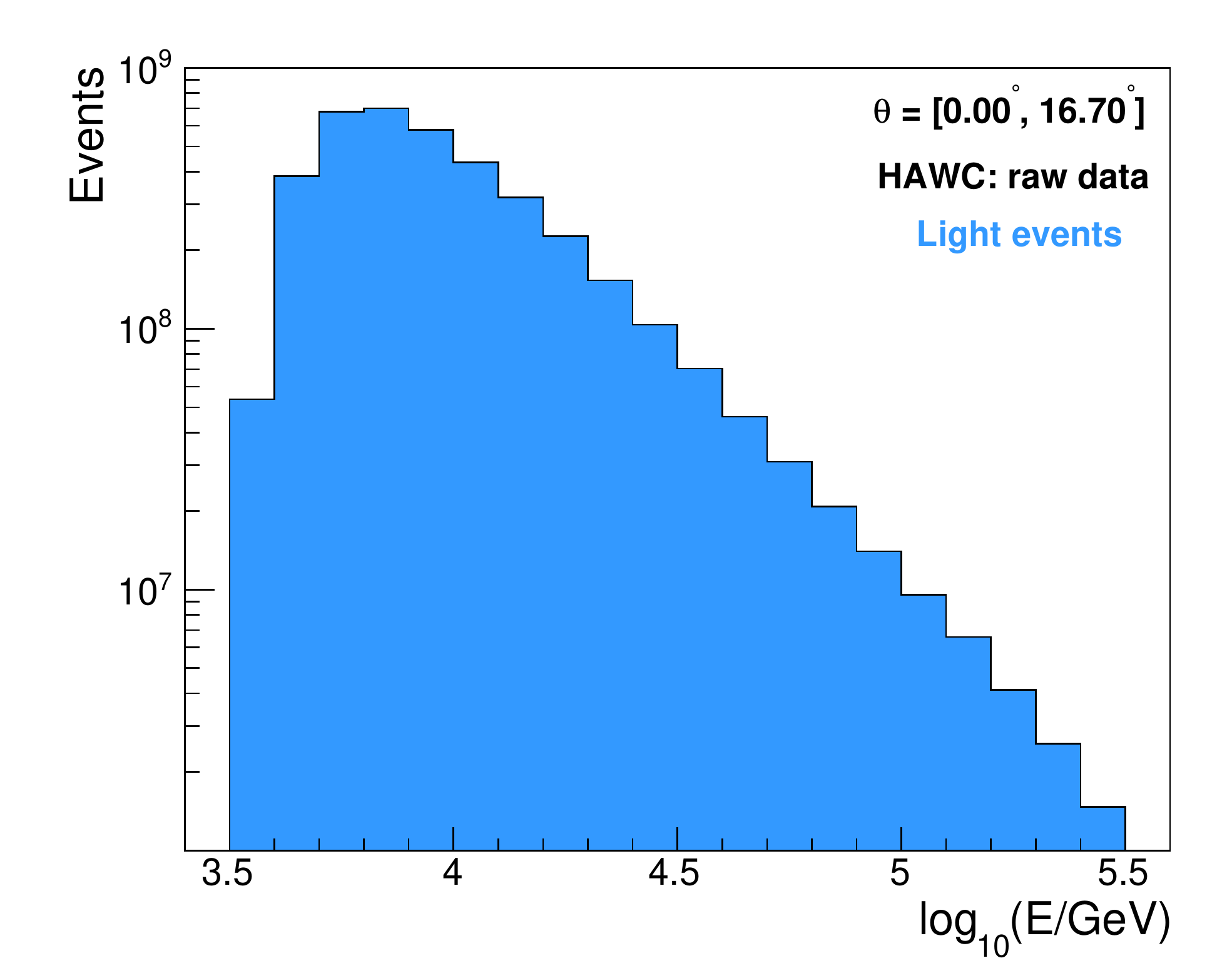}
    \caption{ \textit{Left panel}:  Mean values of the expected age parameter for vertical EAS (circles)
    against the estimated
    shower energy for different primaries and for our nominal composition model 
    (labelled as \textit{all}, open squares). 
    From above to below, the curves with circles correspond
    to Fe, Si, Mg, Ne, O, C, He and H, respectively.
    The HAWC data is also shown for comparison (solid
    squares). Errors on 
    the mean are shown. The segmented line represents the $s_{He-C}$ cut.
    \textit{Right panel}: The histogram of 
    reconstructed energy for the measured 
    sub-sample of light events
    obtained with the cut on the age parameter.}
  	\label{agevse}
\end{figure}

  \section{Reconstruction of the spectrum}
   \vspace{-0.5pc}
  The analysis employed in this work to estimate the spectrum of the
 light mass group of cosmic rays  (H plus He nuclei) is simple and it can 
 be summarized in the following way: First, we select a subset of data enriched 
 with the primaries in which we are interested in. Then we build the energy 
 distribution of the sub-sample and correct it for migration effects using an 
 unfolding algorithm. Next, we correct the unfolded distribution for the 
 contamination of heavy nuclei  ($Z \geq 3$) and the efficiency for H+He 
 primaries. Finally, from the result, we reconstruct the corresponding 
 energy spectrum of the light mass group of cosmic rays.
 
   Now, in order to select our light sub-sample from the data, we exploit 
 the sensitivity of the age parameter to 
 the primary composition. The latter is illustrated, in fig.~\ref{agevse}, left, where the 
 expected mean value of $s$ is presented as a function of the reconstructed 
 energy for different cosmic ray species. The above plot shows that the age parameter 
 increases with the mass of the primary nuclei and, in addition, decreases with 
 $E$. This can be understood from the fact that light primaries and high energy 
 cosmic rays produce air showers with $X_{max}$ closer to the ground and hence 
 with steeper lateral distributions at detection level. To perform the selection, 
 we apply a cut, $s_{He-C}$, on the data, located between the predicted curves for 
 He and C (see fig.~\ref{agevse}, left) in such a way that if the events satisfy 
 $s < s_{He-C}$, then they are classified into the light mass group of cosmic rays,
 otherwise they are considered as a part of the heavy component. 
 Using the above cut on the selected data, we are left with
 a sub-sample of $3.8 \times 10^{9}$ events. 
 In general,
 using MC simulations for our nominal composition model, we found out that the expected
 retention fraction of protons and helium nuclei in the light sub-sample is  
 $\gtrsim 60$\%, while its purity is $\gtrsim 90 \%$.

\begin{figure}[!t]
     \centering
     \footnotesize
     \includegraphics[width=75mm]{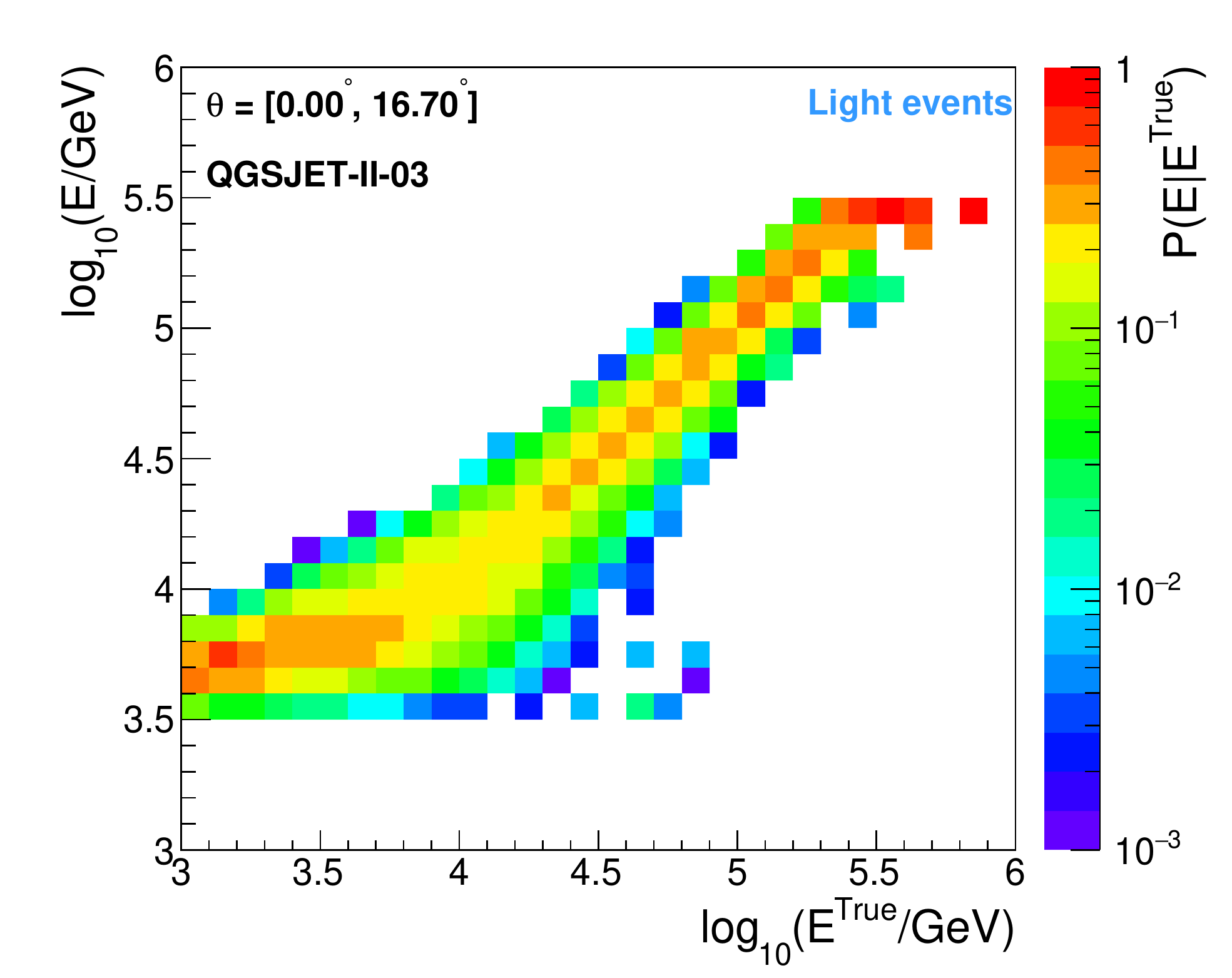}
     \includegraphics[width=75mm]{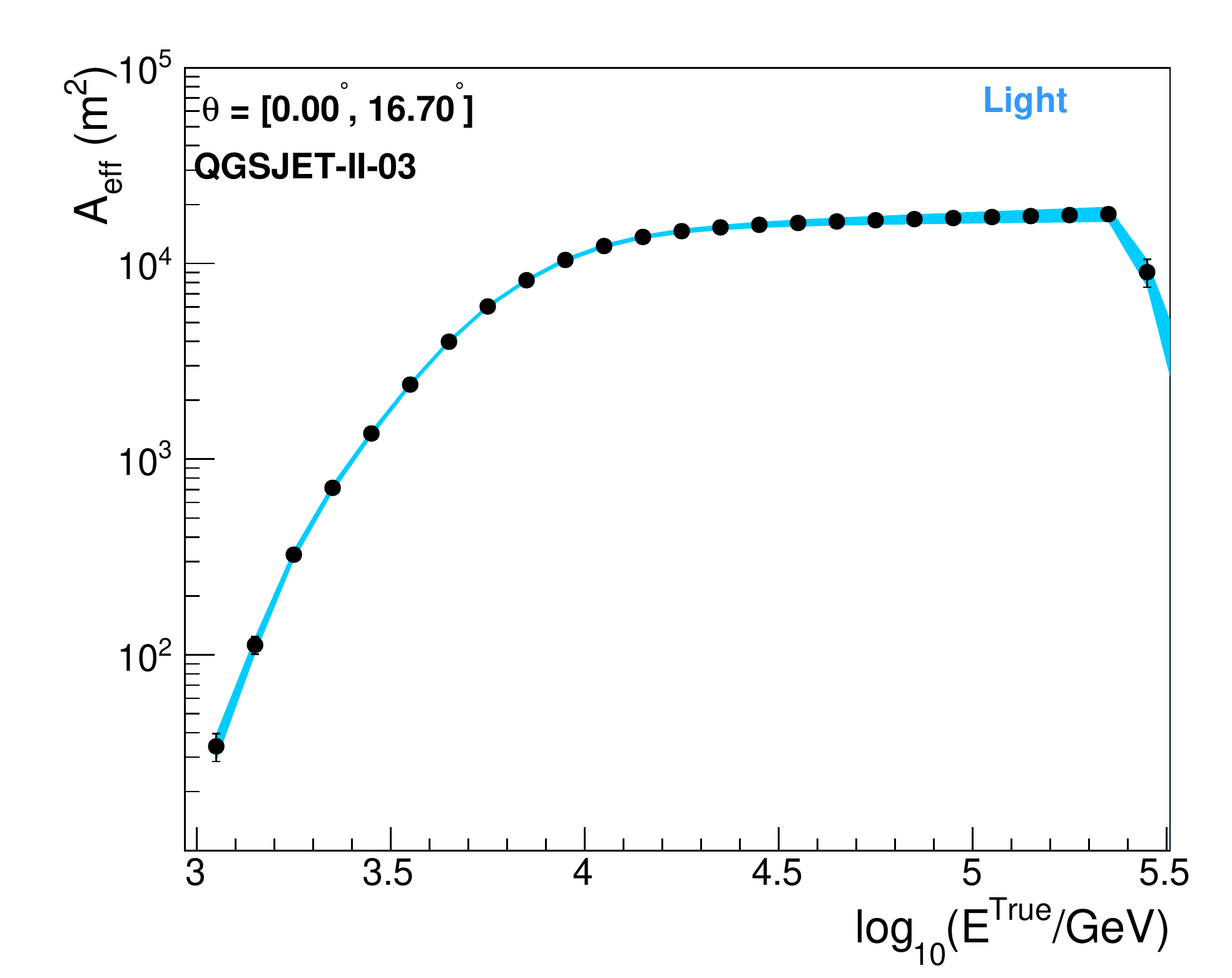} 

     \caption{ \textit{Left panel}: The energy response matrix  for the light sub-sample
            calculated from  our nominal composition model.
            The vertical axis represent the reconstructed energy, and the horizontal
            axis, the true EAS energy. The color code indicates the probability
            $P(E|E^{True})$, which takes into
            account migration effects. 
            \textit{Right panel}: Effective area used for the calculation of the 
     		spectra for the H plus He mass group of cosmic rays. The curve was derived from
     		MC simulations using our nominal composition model. 
    }
  	\label{eff}
    \vspace{-0.5pc}
\end{figure}

  After selecting our light sub-sample, we then build its corresponding energy 
  histogram, $N_{light}^{raw}(E)$, using a bin size of $\Delta \log_{10}(E/\mbox{GeV}) = 0.1$ 
  (see fig.~\ref{agevse}, right). Then we apply the Bayes unfolding procedure \cite{agostini} to 
  correct this distribution for migration effects employing the response matrix, 
  $P(E|E^{True})$ (c.f. fig.~\ref{eff}, left), which was derived for the light subset using 
  MC simulations of our nominal composition model. Using the unfolded histogram
  $N_{light}^{Unf}(E^{True})$, the energy spectrum was calculated according to the 
  following formula:
   \vspace{-0.5pc}
   \begin{equation}
 	\Phi(E) = \frac{N_{light}^{Unf}(E^{True})}{\Delta E^{True} \mathcal{E}(E^{True})},
 	\label{eq1}
 \end{equation}
 where $\Delta E^{True}$ is the width of the energy bin center at $E^{True}$ and 
 $\mathcal{E}(E^{True})$ is the exposure, which is defined as:
  \vspace{-0.5pc}
    \begin{equation}
 	 \mathcal{E}(E^{True}) =  A_{eff}(E^{True}) \cdot \Delta t_{eff} \cdot \Delta \Omega.
 	\label{eq2}
 \end{equation}

 Here, $\Delta t_{eff}$ is the effective time of observation, $d \Omega$ is the differential 
 solid angle and $A_{eff}(E^{True})$ is the effective area, which is defined as
 \begin{equation}
    \vspace{-0.5pc}
 	 A_{eff}(E^{True}) = f_{corr} (E^{True}) \cdot A^{H+He}_{eff}(E^{True})
 	\label{eq3}
 \end{equation}
 with $A^{H+He}_{eff}$, the effective area of protons and helium nuclei in the light sub-sample 
 and $f_{corr}$, a factor that corrects the spectrum for the presence of heavy elements in 
 our data set with light events. Both $A^{H+He}_{eff}$ and $f_{corr}$  were estimated using MC
 simulations. In particular, the effective area was estimated from the expression below \cite{HAWCcrspec}:
  \begin{equation}
 	 A^{H+He}_{eff}(E^{True}) = A_{thrown} \cdot \frac{\cos \theta_{max} + \cos \theta_{min}}{2} 
 	 \cdot \epsilon^{H+He}(E^{True}),
 	\label{eq4}
 \end{equation}
 where $A_{thrown}$ is the total throwing area of the MC events, $\theta_{min}$ and $\theta_{max}$ 
 are the limits of the zenith angle interval considered in this study, and  $\epsilon^{H+He}$ is the 
 efficiency for detecting an hydrogen/helium-induced EAS and classifying it as a 
 part of our light sub-sample. Meanwhile,   $f_{corr}$ was calculated as the inverse of the expected 
 fraction of H and He primaries in this sub-sample. The final result for $ A_{eff}$ is shown in
 fig.~\ref{eff}, right, as a function of the primary 
 energy. We see, from this plot that the maximum efficiency region is found between 
 $\log_{10}(E/\mbox{GeV}) = 4$ and $5.3$. The drop at higher energies in the effective
 area is produced because we are running out of statistics there due to our selection cut 
 at high energies. In the region of maximum efficiency, $f_{corr}$ decreases with energy from 
 $1.13$  up to $1.06$.

\begin{figure}[!t]
     \centering
     \footnotesize
     \includegraphics[width=75mm]{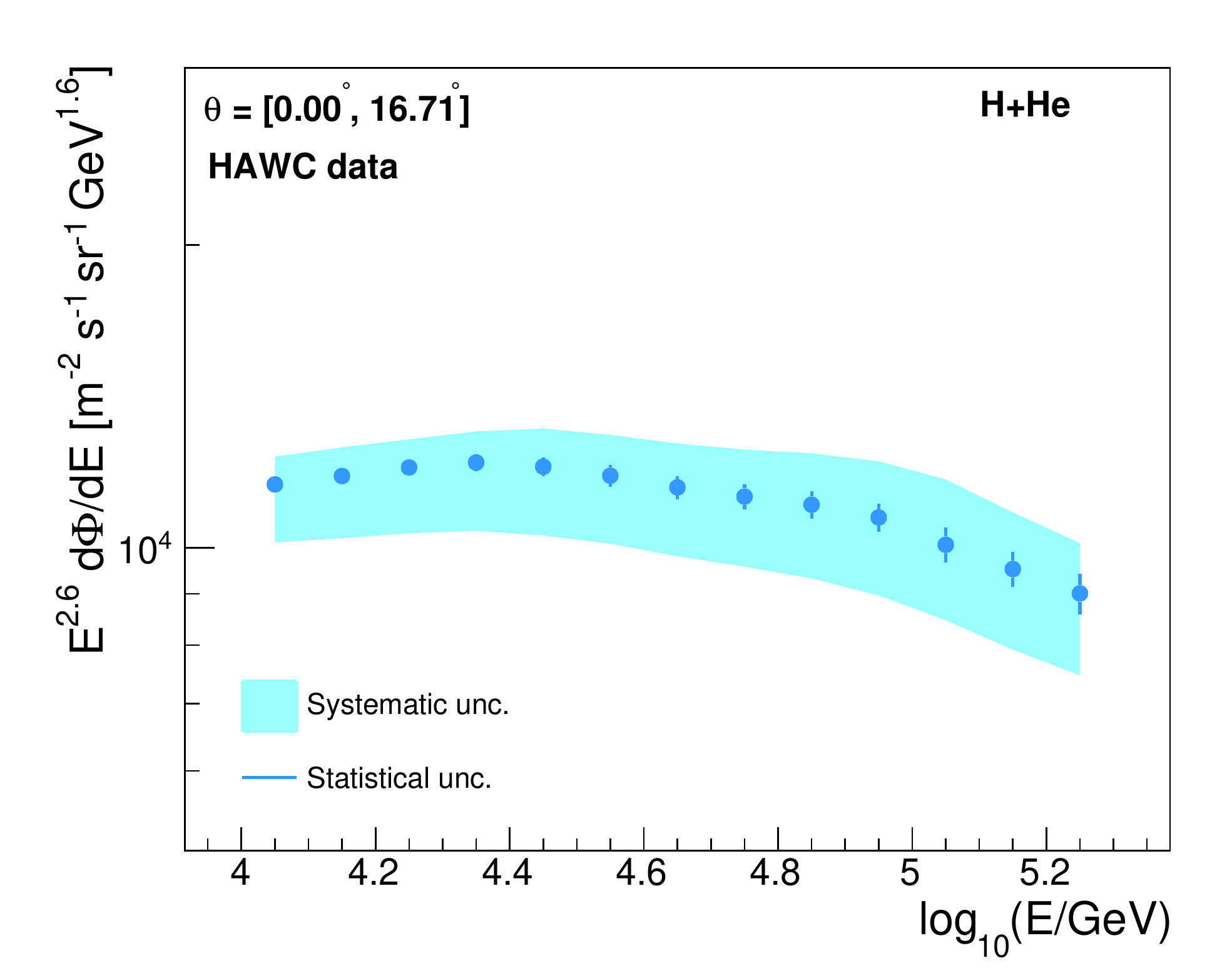}
     \includegraphics[width=75mm]{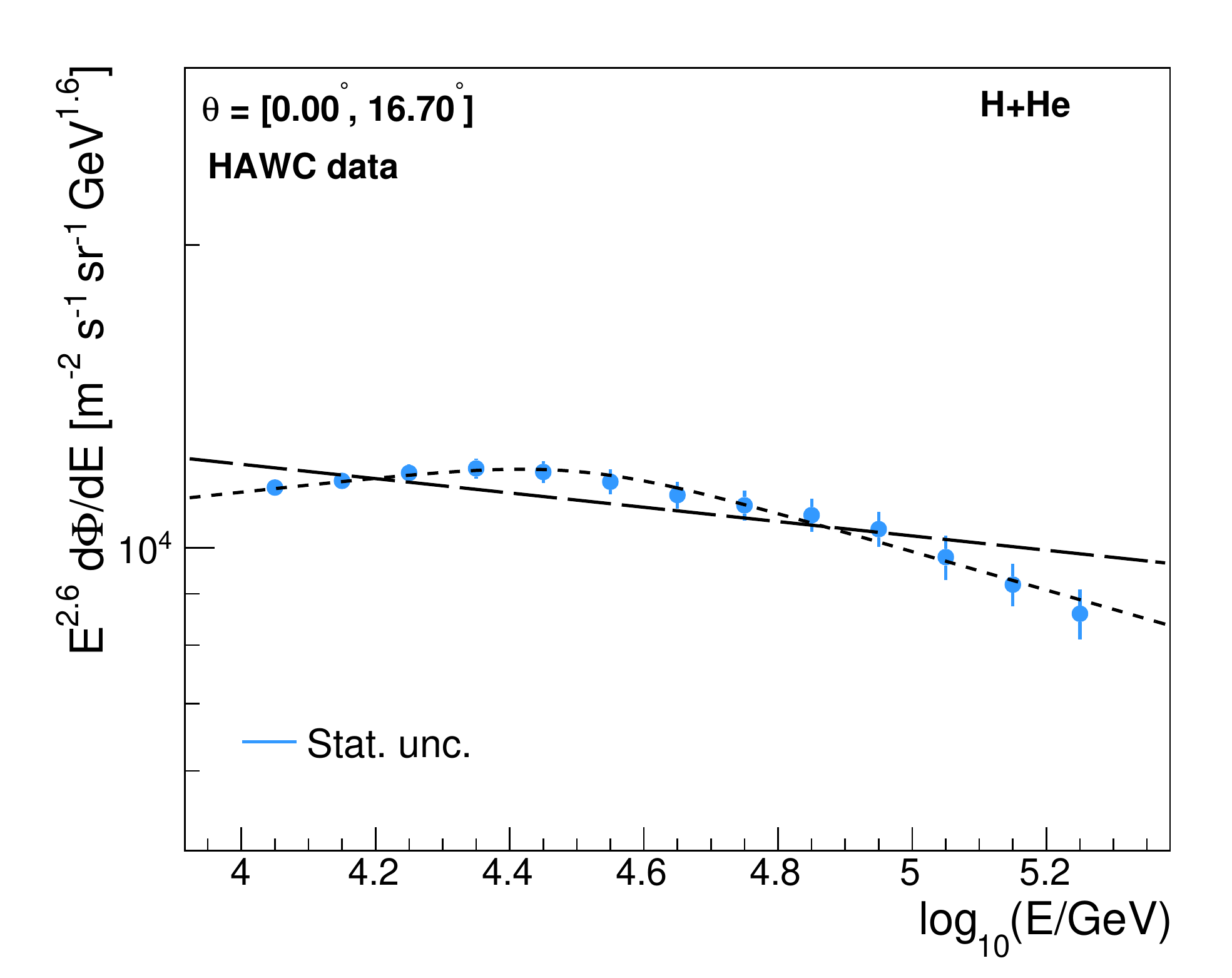} 
     \caption{ \textit{Left panel}: The energy spectrum, multiplied by $E^{2.6}$, for the proton plus helium mass group of cosmic rays as measured with HAWC. The error
     bars represent statistical errors, while
     the error band, systematic uncertainties.
      \textit{Right panel}: The results of the fit to the energy spectrum using a single power-law
      formula (long dashed line) and a double power-law function  (short dashed line) as described in 
      the text. Measured data is represented by
      data points. Error bars are statistical errors.
    }
  	\label{spectrum}
    \vspace{-0.5pc}
\end{figure}

  Finally, the estimated energy spectrum of H plus He is presented in fig.~\ref{spectrum}, left, in the region of 
  full efficiency. Here, the error bars represent the statistical errors and the error band, the 
  systematic uncertainties. The first one varies between $1 \%$ and $5 \%$, while the second one, 
  from $12.2 \%$ to $17 \%$. Statistical errors come from the data size of the sample and the 
  limited statistics of the MC data used to estimate the response matrix. They were calculated 
  according to \cite{Ayde}. Systematic errors include contributions from uncertainties in the 
  primary composition (we used different composition models to reconstruct the spectrum), the 
  calculation of the effective area, the position of the age cut (it was moved between the lines 
  for He and C), the unfolding method (Gold's unfolding \cite{Gold} was also used and the initial 
  spectrum as well as the regularization procedure were changed), the bin size, and the 
  quantum efficiency/resolution of the PMTs.
   
   The dominant source of systematic error in the procedure is the composition dependence of the 
   response matrix and the effective area. The relative error of this systematic source varies 
   in the range  $11 \% - 15 \%$.  In order to evaluate this systematic error, we have estimated 
   both $P(E|E_{True})$ and $A_{eff}$ using different models with distinct elemental abundances 
   as predicted by the Polygonato model \cite{horandel}, and from fits to measurements from 
   ATIC-2 \cite{atic}, MUBEE \cite{mubee} and JACEE \cite{jacee}, and then we have repeated the 
   reconstruction procedure of the spectrum for each case. The variation among the different 
   results including that one obtained using our nominal composition model was cited as the 
   systematic error due to composition.

   \section{Discussion}
      \vspace{-0.5pc}
   \begin{figure}[!t]
     \centering
     \footnotesize
     \includegraphics[width=150mm]{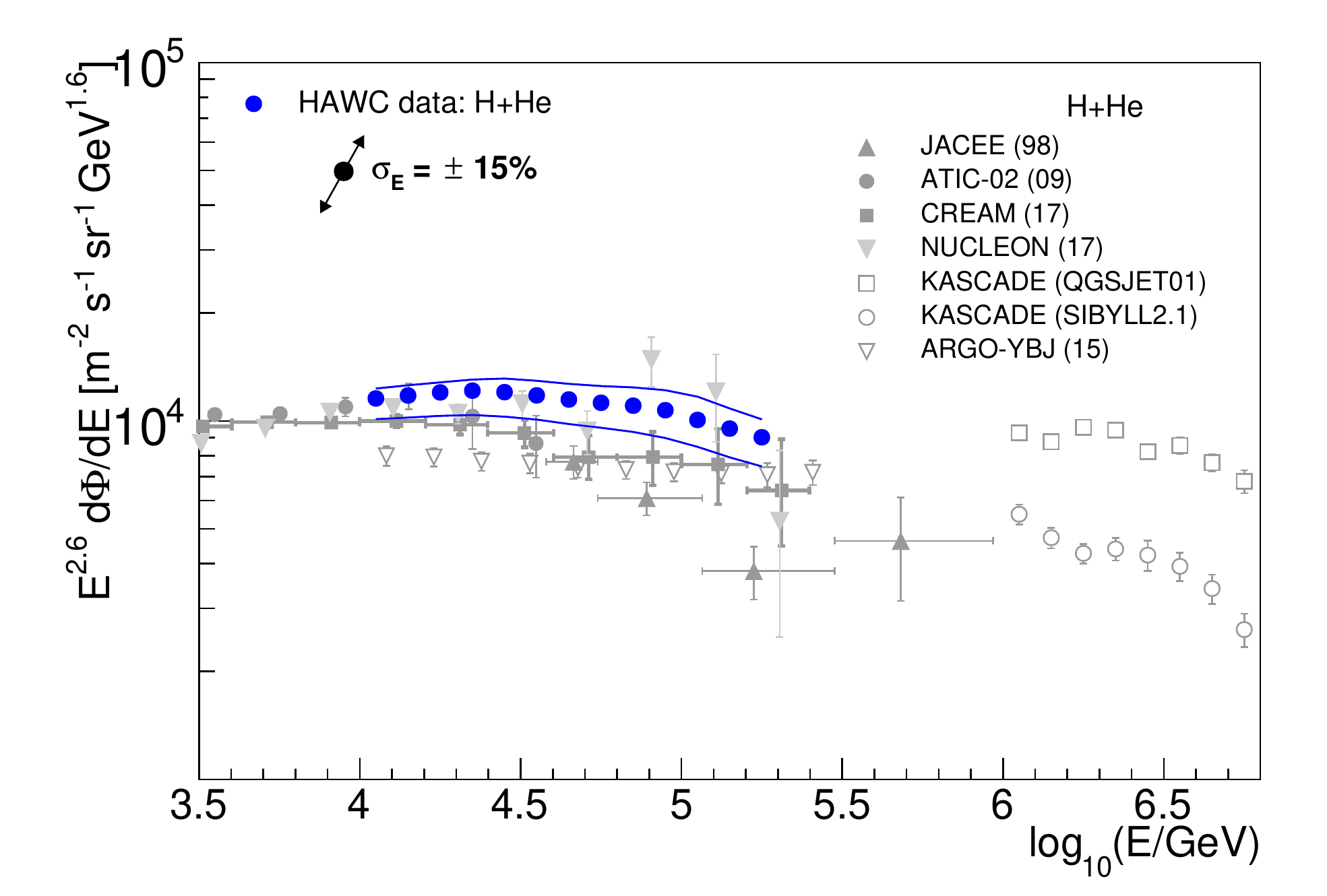}
     \caption{ \textit{Left panel}: The HAWC
     energy spectrum of protons plus helium nuclei (big blue circular points) is compared
     with measurements from direct (ATIC-2 \cite{atic}, MUBEE \cite{mubee} and NUCLEON \cite{nucleon}) and indirect (JACEE \cite{jacee} and KASCADE \cite{kascade})
     experiments. In the HAWC data, statistical errors are smaller than the marker size. On the
     other hand, the error band represents the corresponding systematic uncertainty.}
  	\label{spectrumotherexp}
    \vspace{-0.5pc}
\end{figure}
   
   From figure  fig.~\ref{spectrum}, left, we can observe that the spectrum for protons and helium nuclei seems to deviate from a single power-law behaviour. A $\chi^2$ fit with a single power-law expression,
   \begin{equation}
        \Phi (E)= \Phi_0 E^{\gamma_1},
    	\label{eq5}
   \end{equation}
   taking into account the correlations among the unfolded data points according to \cite{pdg2017}, 
   gives $\gamma_1 = -2.67 \pm 0.01$ and $\log_{10}(\Phi_0/\mbox{m}^{-2} \mbox{s}^{-1} \mbox{sr}^{-1} \mbox{GeV}^{-1}) = 4.37 \pm 0.04$ with $\chi^2 = 30.75$ for $N_{dof} = 11$ degrees of freedom. On 
   the other hand, a $\chi^2$ fit with a double power-law formula, 
   \begin{equation}
 	 \Phi (E)= \Phi_0 E^{\gamma_1} \left[1 + \left(\frac{E}{E_0}\right)^\xi \right]^{(\gamma_2 - \gamma_1)/\xi},
 	\label{eq6}
   \end{equation}
   provides the result $\gamma_1 = -2.53 \pm 0.05$, $\gamma_2 = -2.79 \pm 0.04$, $\log_{10}(\Phi_0/\mbox{m}^{-2} \mbox{s}^{-1} \mbox{sr}^{-1} \mbox{GeV}^{-1}) = 3.77 \pm 0.23$ and $\log_{10}(E_0/\mbox{GeV}) = 4.50 \pm 0.16$ with $\chi^2 = 1.16$ for $N_{dof} = 8$ degrees of freedom. This result seems to imply the existence of a bending
   in the spectrum around $(3.2^{+1.4}_{-1.0}) \times 10^{4} \, \mbox{GeV}$.   The fitted functions are shown on 
   fig.~\ref{spectrum}, right. In order to find out which hypothesis best describes the data, we used the test statistics  $TS = \Delta \chi^2/ \Delta N_{dof}$. First, by employing MC simulations, we generated different correlated data sets to calculate the distribution of $TS$ under the hypothesis of a single power-law behavior. From this distribution, we found a $p$-value $\leq 4.76 \times 10^{-5}$ of having a $TS$ greater or equal the observed one ($TS_{obs} = 9.86$).
   That implies a $3.90 \, \sigma$ deviation from the scenario with a single power-law, which means that it is unlikely that the measured data is described by a single power-law function.
   
   Finally, in fig.~\ref{spectrumotherexp}, we compare the HAWC spectrum for the light mass group of cosmic rays with the measurements from other experiments. We can see that the HAWC spectrum is above the measurements from
   the JACEE \cite{jacee}, ATIC-2 \cite{atic}, CREAM-III \cite{cream17} and ARGO-YBJ  \cite{Argo15, Argo15b} experiments, but it is in agreement with the NUCLEON results
   \cite{nucleon}. However,  within total errors, at low energies HAWC seems to be
   in agreement with ATIC-2 and CREAM-III. We also note that the shape of the HAWC spectrum  
   follows the same trend of the spectra measured by the above mentioned detectors above 
   $10^{4} \, \mbox{GeV}$, but not that of ARGO-YBJ. Hence, HAWC may support previous observations by ATIC-2, CREAM-III
   and NUCLEON, which suggest the existence of a possible new feature in the spectrum of protons plus
   helium nuclei around a few $10 \, \mbox{TeVs}$.

  \section{Conclusions}
      \vspace{-0.5pc}
    In the present study it was found that the lateral age parameter as measured with HAWC is sensitive to the composition of cosmic rays.
    By using such parameter and guided by MC simulations based on QGSJET-II-03, a sub-sample of HAWC data 
    mainly composed by  H and He nuclei was selected, from which the spectrum of the light component (H+He) of cosmic 
    rays in the energy range from $10 \, \mbox{TeV}$ to $200 \, \mbox{TeV}$ was reconstructed. We have found that 
    the spectrum of the protons plus helium in the cosmic ray flux measured with HAWC is not described by a single power-law function, but it seems to be in agreement with a double power-law behaviour. The fit of the data with the latter expression revels a change of the spectral index of the order of $\Delta \gamma = -0.26 \pm 0.07$ in the H+He spectrum close to $(32^{+14}_{-10}) \, \mbox{TeV}$. Further investigation is in progress to evaluate the impact of the high-energy interaction models in these results.\\
 
    \noindent {\bf Acknowledgements}
    The main list of acknowledgements can be found under the following link: 
    https://www.hawc-observatory.org/collaboration/icrc2019.php.
    In addition, J.C.A.V. wants also to thank the partial support from CONACYT grant A1-S-46288.

\end{document}